\documentclass[a4paper,11pt]{article}
\usepackage{amsmath,amssymb,mathrsfs,bm}
\usepackage{jheppub}

\allowdisplaybreaks

\def\del{\partial }			
\def\< #1 >{\left\langle #1 \right\rangle}
\def\l({\left(}
\def\r){\right)}
\def\<{\left\langle}
\def\>{\right\rangle}
\def\no {\nonumber}
\def\limz {\lim_{z\rightarrow 0}}
 		
\def\B {\mathcal{B}}
\def\K {\mathcal{K}}
\def\lB{\mathcal{B}_\mathrm{lin}}
\def\sB{\mathcal{B}_\mathrm{Sch}}
\def\L {\mathcal{L}}
\def\P {\mathcal{P}}
\def\V {\mathcal{V}}

\makeatother

\begin{document}

\thispagestyle{empty}
\vspace*{-2em}
\begin{flushright}
\end{flushright}
\vspace{3.5cm}
\begin{center}
\Large {\bf
The boundary state for a class of analytic solutions\\ in open string field theory}

\vspace{0.5cm}

\normalsize
 \vspace{1.5cm}
 
Daisuke {\sc Takahashi}\footnote{e-mail address:\ \ 
{\tt d-takahashi@rikkyo.ac.jp}}

\vspace{1cm}

{\it 
Department of Physics, Rikkyo University, 
Tokyo 171-8501, Japan}\\

\vspace{3cm}
{\bf Abstract}
\end{center}
We construct a boundary state for a class of analytic solutions in the Witten's open string field theory. The result is consistent with the property of the zero limit of a propagator's length, which was claimed in \cite{Kiermaier:2008qu}. And we show that our boundary state becomes expected one for the perturbative vacuum solution and the tachyon vacuum solution. We also comment on possible presence of multi-brane solutions and ghost brane solutions from our boundary state.

\newpage

\tableofcontents 

\newpage

\section{Introduction}
Schnabl found the analytic solution in the Witten's open string field theory \cite{Witten:1985cc} for the tachyon condensation \cite{Schnabl:2005gv}. The solution was given both by the Schnabl gauge using $b$ ghost zero mode in the sliver frame and wedge states. After that, various analytic solutions were given \cite{Okawa:2006sn}\cite{Schnabl:2007az}\cite{Erler:2007rh}\cite{Okawa:2007ri}\cite{Kiermaier:2007vu}\cite{Kiermaier:2007ki}. And a class of solutions was found in \cite{Okawa:2006vm}, which was written by wedge based states $K$, $B$ and $c$. These states belong to a subspace of the star algebra which is called the $KBc$ subalgebra. The tachyon solution is given simply by using this subalgebra \cite{Erler:2009uj}. 
 Moreover extended solutions were given in \cite{Kiermaier:2010cf}\cite{Bonora:2010hi}\cite{Bonora:2011ri}\cite{Erler:2011tc}\cite{Maccaferri:2011zz}\cite{Noumi:2011kn}\cite{Bonora:2011ns}, which belong to extended subalgebras. Recently, Murata and Schnabl anticipated that multi-brane solutions exist from an energy calculation in the $KBc$ subalgebra \cite{Murata:2011ex}.
\\

In \cite{Kiermaier:2008qu}, the boundary state $|B_*(\Psi)\rangle$ was constructed for any solution in the open string field theory. This state is BRST invariant. Under the gauge transformation of an open string field, the boundary state changes by a $Q$-exact term. Thus, its contraction with on-shell closed string vertex operator is also gauge invariant. This quantity has the same property that has been shown in \cite{Ellwood:2008jh}. Moreover, under variations of both the propagator to construct the boundary state and its length $s$, the boundary state changes by a $Q$-exact term too. If we choose the Schnabl propagator, the boundary state $|B_*(\Psi)\rangle$ is calculable explicitly. Then, we use the natural $z$ frame that is not usual sliver frame. This map is consistent with identification in a construction of one loop in the annulus frame. In \cite{Kiermaier:2008qu}, choosing the tachyon vacuum solution and the solutions for marginal deformations as the solution, expected boundary states were obtained.
\\

In this paper, we construct a boundary state for the class of analytic solutions. In fact, it becomes expected one for both the perturbative vacuum solution and the tachyon vacuum solution; 
	\begin{align}
        	&|B_*(\Psi_\mathrm{per})\rangle \;= |B\rangle\\
                &|B_*(\Psi_\mathrm{tach})\rangle = 0.
        \end{align}
In the first line, $|B\rangle$ denotes the boundary state when a D-brane exists. The second line means that the D-brane vanishes at the tachyon vacuum. And our boundary state is consistent with the claim of \cite{Kiermaier:2008qu} that for zero limit of $s$ the boundary state coincides with the zeroth and the first order boundary state in the expansion in terms of a solution. Moreover we comment on multi-brane solutions using our boundary state. For $s\rightarrow 0$, the result is identified with \cite{Murata:2011ex}. But the boundary state of finite $s$ is not consistent with \cite{Kiermaier:2008qu}. Thus multi-brane solutions can not be obtained from the form proposed in \cite{Murata:2011ex}.
\\

This paper is organized as follows. In section 2 we review the $KBc$ subalgebra and a solution using it. In section 3 we also review a construction of the boundary state and its properties. In section 4 we construct the boundary state for the class of analytic solutions. And we calculate the boundary state of the perturbative solution and the tachyon vacuum solution for instance. In section 5 we comment on multi-brane solutions and ghost brane solutions (cf.\cite{Okuda:2006fb}). Section 6 is devoted to conclusions and discussions.

\section{Analytic solution in $KBc$ subalgebra}
$KBc$ subalgebra is constructed on the sliver frame which is mapped from the radial frame $\xi$;
	\begin{align}
        	z = g(\xi) = \frac{2}{\pi}\arctan \xi.\label{sliver_frame}
        \end{align}
$z$ denotes the coordinate on the sliver frame. On this frame, the operator $\B$ is a line integral of the $b$ ghost defined by
	\begin{align}
        	\B = \int_{i\infty}^{-i\infty}\frac{dz}{2\pi i}b(z).
        \end{align}
Similarly, the operator $\K$ is defined by
	\begin{align}
        	\K = \int_{i\infty}^{-i\infty}\frac{dz}{2\pi i}T(z),
        \end{align}
where $T(z)$ is the energy momentum tensor.

Using these two operators, the $c$ ghost and the identity state $|I\rangle$, we define states as follows;
  	\begin{align}
        	K &= \K |I\rangle, \\
                B &= \B |I\rangle, \\
                c &= c(\tfrac{1}{2}) |I\rangle,
        \end{align}
where $\frac{1}{2}$ is a position at which the $c$ ghost is inserted on the real axis of the sliver frame. The states $K$, $B$ and $c$ satisfy the following relations;
	\begin{align}
        	[K,B] = 0,\quad \{ B, c \}=1,\quad [K,c]=\del c, \quad B^2 =0,\quad c^2 =0.
        \end{align}
We do not explicitly write $*$ that means a product of string fields. The BRST operator acts on these states in the following way;
	\begin{align}
        	QB=K, \qquad QK = 0, \qquad Qc = cKc.
        \end{align}

The wedge state $W_{\alpha}$ with a width $\alpha \geq 0$ is defined by a BPZ inner product as follows;
	\begin{align}
        	\langle \phi, W_{\alpha} \rangle = \langle g\circ \phi(0)\rangle_{\alpha+1},
        \end{align}
where $\phi(\xi)$ is a generic state in the Fock space and $g\circ \phi(\xi)$ denotes the conformal transformation of $\phi(\xi)$. On the sliver frame, the subscript as in the right hand side means a perimeter of cylinder on which a correlator is defined. The wedge state $W_\alpha$ can be expressed by $K$ as follows;
	\begin{align}
        	W_{\alpha} = e^{\alpha K}.
        \end{align}

The class of analytic solutions is expressed by using these states;
	\begin{align}
        	\Psi = Fc\frac{KB}{1-F^2}cF, \label{solution}
        \end{align}
where $F$ is an arbitrary analytic function of $K$. The solution of this form have been found first by Okawa \cite{Okawa:2006vm}. We assume that $F$ can be written as a kind of Laplace transformation;
	\begin{align}
        	F(K) = \int^\infty_0 dt f(t)e^{tK}.
        \end{align}
So we can regard $F(K)$ as a superposition of wedge states \cite{Erler:2006ww}\footnote{We use the right handed convention for the star product in this paper \cite{Kiermaier:2010cf}\cite{Noumi:2011kn}.}. For example, when we select $\frac{1}{\sqrt{1-K}}$ as $F(K)$, 
	\begin{align}
        	\frac{1}{\sqrt{1-K}} = \int^\infty_0 dt \frac{e^{-t}}{\sqrt{\pi t}}e^{tK},
        \end{align}
we obtain the tachyon vacuum solution \cite{Erler:2009uj}. We also assume that the string field $\frac{K}{1-F^2}$ can be written as a kind of Laplace transformation of $\tilde{f}$. Thus we can write (\ref{solution}) as follows;
	\begin{align}
        	 Fc\frac{KB}{1-F^2}cF &= \int^\infty_0 dt_1 f(t_1) \int^\infty_0 dt_2 \tilde{f}(t_2) \int^\infty_0 dt_3 f(t_3) \Psi_{\mathrm{integrand}}\no\\
                 		      &= \int^\infty_0 dt_1 f(t_1) \int^\infty_0 dt_2 \tilde{f}(t_2) \int^\infty_0 dt_3 f(t_3)e^{t_1K}ce^{t_2K}Bce^{t_3K}.\label{analytic_integrand}
        \end{align}

The BPZ inner product of a test state $\phi$ and (\ref{solution}) is expressed as;
	\begin{align}
        	&\Big\langle \phi ,  Fc\frac{KB}{1-F^2}cF \Big\rangle\no\\
                & = \int^\infty_0 dt_1 f(t_1) \int^\infty_0 dt_2 \tilde{f}(t_2) \int^\infty_0 dt_3 f(t_3)\langle g\circ \phi(0)c(\tfrac{1}{2}+t_1)\B c(\tfrac{1}{2} + t_1 + t_2) \rangle_{1+t_1+t_2+t_3}.\label{op_sol}
        \end{align}
This is useful because we can see both positions of $c$ ghosts and a width of a string field in a correlator on the sliver frame.

\section{Review of the boundary state in open string field theory}
We review the boundary state that has been constructed by Kiermaier, Okawa and Zwiebach \cite{Kiermaier:2008qu}. After that, we consider the boundary state for the string field (\ref{solution}) in the next section. 

\subsection{Construction of the boundary state from open string fields}
Before we construct the boundary state, we introduce a propagator strip for linear $b$-gauges \cite{Kiermaier:2007jg}. Its gauge condition is 
	\begin{align}
        	\lB \Psi = 0, \label{gauge_condition}
        \end{align}
where $\Psi$ is an open string field and the operator $\lB$ is 
	\begin{align}
        	\lB = \oint \frac{d\xi}{2 \pi i}\frac{\mathcal{F}(\xi)}{\mathcal{F}'(\xi)}b(\xi).\label{B_linear}
        \end{align}
Therefore, $\lB$ is the zero mode of the $b$ ghost in a frame specified by $\mathcal{F}(\xi)$ frame. Here, we introduce a $w$ frame related to the $z$ frame (\ref{sliver_frame})\footnote{Of course, the $w$ frame is generally related to the $\mathcal{F}$ frame \cite{Kiermaier:2008qu}\cite{Kiermaier:2008jy} };
	\begin{align}
        	z = \frac{1}{2}e^w.
        \end{align}
Then, the imaginary part of a string midpoint is $\frac{\pi}{2}$ in the $w$ frame. And two string endpoints $\Re z > 0$ or $<0$ correspond to $\Im w = 0$ or $\pi$, respectively. The propagator of linear $b$ gauges is given by $e^{-s\L}$, where
	\begin{align}
        	\L = \{Q, \lB \}.
        \end{align}
In the $w$ frame, $\L$ is the generator of a horizontal translation.

\begin{figure}[t]
		  	\begin{center}
   				\includegraphics[width=70mm]{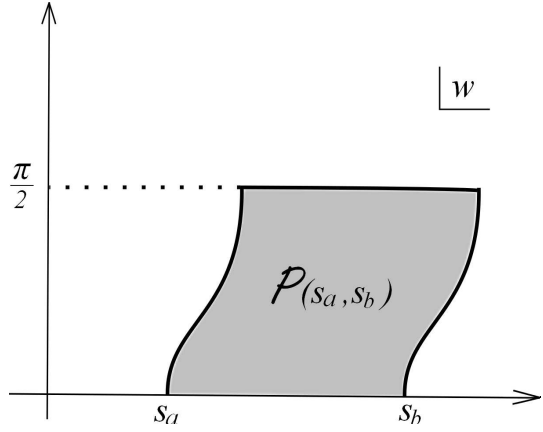}
  			\end{center}
  				\caption{The surface $\P(s_a, s_b)$ is a half-propagator strip. The surface is obtained by cutting a open string world sheet along a median.}
 		 		\label{fig:one}
 	\end{figure}

\begin{figure}[t]
	  		\begin{center}
   				\includegraphics[width=70mm]{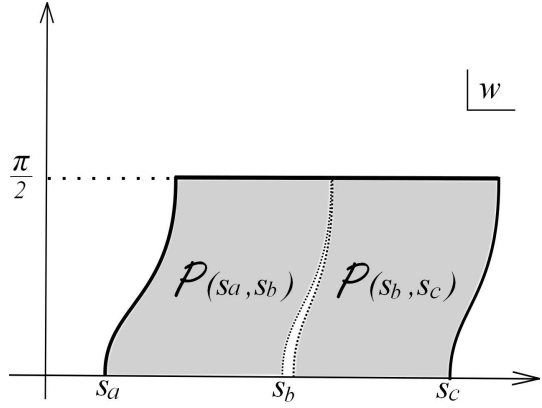}
  			\end{center}
 		 	\caption{The product of surfaces $\P$ with a string field. The string field is inserted into a slit.}
  		 	\label{fig:two}
	\end{figure}

We consider a half-propagator strip $\P(s_a,s_b)$ to construct the boundary state. This strip is obtained by dividing a open string world sheet along a median. See Figure \ref{fig:one}.
More precisely, to construct the half-propagator strip, we use a operator $\L_R(t)$, where $\L_R$ is the right half of $\L$. We define this operator as follows;
	\begin{align}
        	\L_R(t) \equiv \int_t^{\gamma(\frac{\pi}{2}) + t} \left[ \frac{dw}{2 \pi i}T(w) + \frac{d\bar{w}}{2 \pi i}\tilde{T}(\bar{w})  \right],\label{right_L}
        \end{align}
where $\gamma(\theta) = w(\xi = e^{i\theta})$, and the surface $\P(s_a, s_b)$ is expressed by the path-ordered exponential as follows;
	\begin{align}
        	\P(s_a, s_b) = \mathrm{P}\exp\left[ -\int^{s_b}_{s_a}dt\L_R(t) \right],
        \end{align}
where P is the symbol of the path-order product. The star multiplication of these surfaces satisfies
	\begin{align}
        	\P(s_a, s_b)\P(s_b, s_c) = \P(s_a, s_c).
        \end{align}
A surface glued to a string field $A$ is expressed as $\P(s_a, s_b)A\P(s_b, s_c)$. This surface is expressed geometrically as Figure \ref{fig:two}.
        
A one point function where a closed string vertex operator $\phi_c$ is inserted in the origin on a unit disk is written by using a boundary state $|B\rangle$;
	\begin{align}
        	\langle (c_0 - \tilde{c}_0) \phi_c(0) \rangle_{\mathrm{disk}} = \langle B| (c_0 - \tilde{c}_0) | \phi_c \rangle,
        \end{align}
where the operator $c_0 - \tilde{c}_0$ is associated with a conformal Killing vector on the disk. This correlator can be seen as an inner product of $\langle B|e^{-\frac{\pi^2}{s}(L_0 + \tilde{L}_0)}$ and $e^{\frac{\pi^2}{s}(L_0 + \tilde{L}_0)} (c_0 - \tilde{c}_0) | \phi_c \rangle$, in which we cut the disk along a circle of radius $e^{-\frac{\pi^2}{s}}$ in a $\zeta=\exp\l(\frac{2\pi i}{s} w \r)$ frameG
	\begin{align}
        	\langle B| (c_0 - \tilde{c}_0) | \phi_c \rangle = \langle B|e^{-\frac{\pi^2}{s}(L_0 + \tilde{L}_0)}e^{\frac{\pi^2}{s}(L_0 + \tilde{L}_0)} (c_0 - \tilde{c}_0) | \phi_c \rangle.
        \end{align}
So, in the following, we can write the relation between the surface $\P(0,s)$ and the boundary state $|B\rangle$;
	\begin{align}
        	\oint_s \P(0,s) = e^{-\frac{\pi^2}{s}(L_0 + \tilde{L}_0)}|B\rangle,
        \end{align}
where the operation $\oint_s$ means identification of a left boundary and a right boundary of the surface $\P(0,s)$. Therefore we obtain the boundary state as follows;
	\begin{align}
        	|B\rangle = e^{\frac{\pi^2}{s}(L_0 + \tilde{L}_0)}\oint_s \P(0,s).
        \end{align}

Moreover, we define the boundary state around the classical solution $\Psi$ of the equation of motion $Q\Psi + \Psi^2 = 0$. To do this, we replace $\L = \{ Q , \lB\}$ with $\{ Q_* , \lB\}$, where $Q_*$ is the BRST operator around the new background;
	\begin{align}
        	Q_*A = QA +\Psi A - (-)^AA\Psi.
        \end{align}
When $\{Q_*,\lB\}$ acts on a string field $A$,
	\begin{align}
        	 \{Q_*,\lB\} A = \left[\L_R A + (-)^A(\B_R A)\Psi - (-)^A\B_R(A \Psi)  \right] + \left[ \L_L A + \Psi (\B_L A) + \B_L (\Psi A)  \right],
        \end{align}
where 
	\begin{align}
        	\B_R(t)  \equiv \int_t^{\gamma(\frac{\pi}{2}) + t} \left[ \frac{dw}{2 \pi i}b(w) + \frac{d\bar{w}}{2 \pi i}\tilde{b}(\bar{w})  \right]\label{right_B}
        \end{align}
is the right part of $\lB$ and $\B_L$ is the left part of it; $\lB = \B_R + \B_L$. Therefore we replace $\L_R(t)$ with $\L_R(t) + \{ \B_R(t), \Psi \}$ for $\P(s_a,s_b)$;
	\begin{align}
        	\P(s_a,s_b) \rightarrow \P_*(s_a,s_b) \equiv \mathrm{P}\exp\left[ -\int^{s_b}_{s_a}dt \left( \L_R(t) + \{ \B_R(t),\Psi \} \right) \right].
        \end{align}
 Using this new surface $\P_*(s_a,s_b)$, we express the boundary state $| B_*(\Psi) \rangle$ around the new background in analogy with $|B\rangle$;
 	\begin{align}
        	| B_*(\Psi) \rangle = e^{\frac{\pi^2}{s}(L_0 + \tilde{L}_0)}\oint_s \P_*(0,s).
        \end{align}
We can write the boundary state by expanding $\P_*(0,s)$ in powers of the classical solution;
	\begin{align}
        	&|B_*(\Psi) \rangle\no\\
                =& \sum_{k=0}^{\infty} | B_*^{(k)}(\Psi)\rangle \no \\
                =& \sum_{k=0}^{\infty} (-)^ke^{\frac{\pi^2}{s}(L_0+\tilde{L}_0)}\oint_s \int_0^s ds_1 \cdots \int_{s_{i-1}}^s ds_i \cdots \int_{s_{k-1}}^s ds_k \P(0,s_1)\{ \B_R(s_1) ,\Psi  \}\P(s_1,s_2)\no\\
                &\qquad\qquad\qquad \cdots \P(s_{i-1},s_i)\{ \B_R(s_i) ,\Psi  \}\P(s_i,s_{i+1})\cdots \P(s_{k-1},s_k)\{ \B_R(s_k) ,\Psi  \}\P(s_k,s).\label{general_boundary_state}
        \end{align}
Clearly, $|B_*^{(0)}(\Psi)\rangle = |B\rangle$. We can actually recognize $|B_*(\Psi)\rangle$ as the boundary state because it has expected properties. They are given in the next subsection.
\subsection{Properties of the boundary state}
We introduce various properties of $|B_*(\Psi)\rangle$ without proofs, which are in the paper \cite{Kiermaier:2008qu}.
\\\\
\textbf{Fundamental properties}\\
We can recognize $|B_*(\Psi)\rangle$ as the boundary state because it satisfies three properties. They are
	\begin{align}
        	Q|B_*(\Psi)\rangle = 0,\qquad (b_0 - \tilde{b}_0)|B_*(\Psi)\rangle = 0,\qquad (L_0 - \tilde{L}_0)|B_*(\Psi)\rangle = 0.
        \end{align}
The equation of motion of the open string field theory is used to prove the BRST invariance of $|B_*(\Psi)\rangle$
\\\\
\textbf{Property under open string gauge transformation}\\
An open string infinitesimal gauge transformation is 
	\begin{align}
        	\delta_\chi \Psi = Q\chi + [\Psi,\chi].
        \end{align}
When we gauge transform a string field $\Psi$ in $|B_*(\Psi)\rangle$ given in (\ref{general_boundary_state}), the boundary state becomes
	\begin{align}
        	\delta_\chi |B_*(\Psi)\rangle = Q-\mathrm{exact}.\label{gauge_property}
        \end{align}
\\\\
\textbf{Variation of the propagator}\\
We consider an infinitesimal change of gauge condition (\ref{gauge_condition}) for the propagator. To do that, we change two operators;
	\begin{align}
        	\B_R(t) \rightarrow \B_R(t) + \delta\B_R(t),\qquad \L_R(t) \rightarrow \L_R(t) + \{\mathcal{Q}_R(t), \delta \B_R\},
        \end{align}
where $\mathcal{Q}_R$ is, using the BRST current $j_B$,
	\begin{align}
        	\mathcal{Q}_R = \int^{\gamma(\frac{\pi}{2})+ t}_t \left[ \frac{dw}{2 \pi i}j_B(w) - \frac{d\bar{w}}{2 \pi i}\tilde{j}_B(\bar{w}) \right].
        \end{align}
They produce a change of the boundary state as follows;
	\begin{align}
        	\delta|B_*(\Psi)\rangle = Q-\mathrm{exact}.\label{propagator_pro}
        \end{align}
\\\\
\textbf{Variation of $\bm{s}$}\\
The boundary state (\ref{general_boundary_state}) depends on $s$.  An infinitesimal variation with respect to $s$ of the boundary state is also $Q$-exact;
	\begin{align}
        	\del_s |B_*(\Psi)\rangle = Q - \mathrm{exact}.\label{s_property}
        \end{align}
The equation of motion is used to prove (\ref{gauge_property})(\ref{propagator_pro})(\ref{s_property}) too.
Therefore the inner product of on-shell closed string vertex $\mathcal{V}$ and $|B_*(\Psi)\rangle$ is invariant under the gauge transformation, the variation of the propagator and that of $s$.
\\\\
\textbf{$\bm{s\rightarrow 0}$ limit}\\
For $s \rightarrow 0 $, the boundary state $|B_*(\Psi)\rangle$ becomes only its zeroth and first order;
	\begin{align}
        	|B_*(\Psi)\rangle = |B\rangle + |B_*^{(1)}(\Psi) \rangle.
        \end{align}
Although we do not have a rigorous proof of this, this holds in specific examples \cite{Kiermaier:2010cf} (and we will see ones in the next section).
\\\\
\textbf{Gauge invariant observable $\bm{W(\V,\Psi)}$}\\
For $s \rightarrow 0$, an inner product $\langle \V|(c_0 - \tilde{c}_0)|B_*(\Psi)\rangle$ is written as follows;
	\begin{align}
        	\lim_{s\rightarrow 0}\langle \V|(c_0 - \tilde{c}_0)|B_*(\Psi)\rangle = \langle \V|(c_0 - \tilde{c}_0)|B\rangle + \langle \V|(c_0 - \tilde{c}_0)|B_*^{(1)}(\Psi)\rangle.
        \end{align}
This relation implies the identification of the gauge invariant observable $W(\V,\Psi)$ and disk amplitudes \cite{Ellwood:2008jh};
	\begin{align}
        	\lim_{s\rightarrow 0}\langle \V|(c_0 - \tilde{c}_0)|B_*(\Psi)\rangle - \langle \V|(c_0 - \tilde{c}_0)|B\rangle =  -4\pi iW(\V,\Psi).
        \end{align}
 Moreover, for the variation of $s$, the boundary state only changes by a $Q$-exact term. Thus, because $\V$ is on-shell, 
 	\begin{align}
        	\langle \V|(c_0 - \tilde{c}_0)|B_*(\Psi)\rangle - \langle \V|(c_0 - \tilde{c}_0)|B\rangle =  -4\pi iW(\V,\Psi) \qquad \mathrm {for\;\;any}\;\;s.
        \end{align}
\subsection{Boundary state for Schnabl gauge}
We can calculate $|B_*(\Psi)\rangle$ simply by using Schnabl gauge. The zero mode of $b$ ghost in the gauge fixing condition (\ref{B_linear}) is 
	\begin{align}
        	\sB = \oint \frac{d\xi}{ 2 \pi i}\frac{g(\xi)}{g'(\xi)}b(\xi),
        \end{align}
where we take $\mathcal{F(\xi)}=g(\xi)$ given in (\ref{sliver_frame}). In the $z$ frame, two operators(\ref{right_L})(\ref{right_B}) are defined as follows;
	\begin{align}
        	\L_R(t) = \int_{C(t)}\frac{dz}{2 \pi i}zT(z),\qquad \B_R(t) = \int_{C(t)}\frac{dz}{2 \pi i}zb(z),
        \end{align}
where the contour $C(t)$ runs from $\frac{e^t}{2} -i\infty $ to $\frac{e^t}{2} + i\infty $. The surface $\P(0,s)$ is located in the region
	\begin{align}
        	\frac{1}{2} \leq \Re z \leq \frac{1}{2}e^s. 
        \end{align}
Thus, in this frame, the operation $\oint_s$ means identification $z \sim e^s z$. This is compatible with the $\zeta$ frame. When we insert a string field $A_{\alpha_1}$ that has a width $\alpha_1$ into $s_1$ in the $w$ frame, the left part of $A_{\alpha_1}$ is glued to $\frac{1}{2}e^{s_1}$ and the right part is to $e^{s_1}\alpha_1 + \frac{1}{2}e^{s_1}$ in the $z$ frame. Generally, after $i-1$ string fields are inserted, the left part of $A_{\alpha_i}$ is glued to $\sum_{j=1}^{i-1}e^{s_j}\alpha_j + \frac{1}{2}e^{s_i} $ and the right part is to $\sum_{j=1}^{i}e^{s_j}\alpha_j + \frac{1}{2}e^{s_i}$. Therefore, when $k$ string fields are inserted into $\P(0,s)$, the region is
	\begin{align}
        	\frac{1}{2} \leq \Re z \leq \sum_{j=1}^{k} e^j \alpha_j + \frac{1}{2}e^s
        \end{align}
in the $z$ frame. But this is not identification $z \sim e^s z$. Thus we shift the entire surface horizontally by
	\begin{align}
        	a_0 = \frac{1}{e^s - 1}\sum_{j=1}^k e^{s_j}\alpha_j,
        \end{align}
and this region becomes
	\begin{align}
        	\frac{1}{2} + a_0 \leq \Re z \leq e^s\l( \frac{1}{2} + a_0 \r).
        \end{align}
The identification $z \sim e^{s}z$ is restored. This translated frame is called the natural $z$ frame. The map from $z^{(i)}$ of the wedge surface on which $A_{\alpha_i}$ is defined to the natural $z$ frame is given by
	\begin{align}
        	z = \ell_i + e^{s_i}z^{(i)},
        \end{align}
where 
	\begin{align}
        	\ell_i = \sum_{j=1}^{i-1}\alpha_j e^{s_j} + a_0,\qquad \ell_1 = a_0.
        \end{align}

At the end of the subsection, we introduce two useful formulae for the calculation of $|B_*(\Psi)\rangle$;
	\begin{align}
        	-\{ \B_R(s_i), A_{\alpha_i} \}& \rightarrow \oint \frac{dz}{2 \pi i}(z - \ell_i)b(z)[\cdots] - e^{s_i}\alpha_i [\cdots]\B \label{formula1}\\
                \B[\cdots] &= \frac{e^s}{e^s - 1}\oint \frac{dz}{2 \pi i}b(z)[\cdots]. \label{formula2}
        \end{align}
In the first formula, $[\cdots]$ means operator insertions which the string field $A_{\alpha_i}$ has. And for the second formula, we can only use this when other operator insertions are not in the boundary state.

\section{The boundary state for a class of analytic solutions}
We consider the boundary state $|B_*(\Psi)\rangle$ for the case of  $\Psi$ being the analytic solution (\ref{solution}). For this purpose, we should carefully find positions of $c$ ghosts and a width of string fields. Thus it is convenient to see a description like (\ref{op_sol}). First, we calculate the $k$-th order boundary state $|B_*^{(k)}(\Psi)\rangle$ in (\ref{general_boundary_state}). In fact, what we calculate is
	\begin{align}
        	\prod_{i=1}^{k} [-\{ \mathcal{B}_R(s_i) , \Psi^{(i)} \}].
        \end{align}
The $i$-th string field in the description like (\ref{analytic_integrand})(\ref{op_sol}) is written as follows;
	\begin{align}
                &\langle \phi, \Psi^{(i)} \rangle\no\\
                &= \int_0^{\infty}d\alpha_i f(\alpha_i)\int_0^{\infty}d\beta_i \tilde f(\beta_i)\int_0^{\infty}d\gamma_i f(\gamma_i) \Big\langle \phi, \Psi^{(i)}_{\mathrm {integrand}} \Big\rangle\no\\
                &= \int_0^{\infty}d\alpha_i f(\alpha_i)\int_0^{\infty}d\beta_i \tilde f(\beta_i)\int_0^{\infty}d\gamma_i f(\gamma_i)  \Big\langle g\circ \phi(0) c\left(\tfrac{1}{2} + \alpha_i \right) \B c\l( \tfrac{1}{2} + \alpha_i + \beta_i  \r) \Big\rangle_{1+ \alpha_i + \beta_i + \gamma_i}.\label{i-th_string_field}
        \end{align}
In the natural $z$ frame, operators inserting in $i$-th string field are conformal transformed as $c(\tfrac{1}{2}+t) \rightarrow e^{-s_i}c\l( e^{s_i}\l( \tfrac{1}{2} + t \r) + \ell_i\r)$ and $\B \rightarrow e^{s_i}\B$. Therefore, from the width and positions of the $c$ ghosts of the $i$-th string field in (\ref{i-th_string_field}), using (\ref{formula1}),
	\begin{align}
        	&\prod_{i=1}^{k} [-\{ \mathcal{B}_R(s_i) , \Psi^{(i)}_{\mathrm{integrand}} \}]\no\\
                &\rightarrow\prod_{i=1}^{k}\Big\{ e^{-s_i}\oint \frac{dz}{2 \pi i}(z- \ell_i)b(z)c\l( e^{s_i}\l( \tfrac{1}{2} + \alpha_i \r) + \ell_i \r)\B c(e^{s_i}(\tfrac{1}{2} + \alpha_i + \beta_i) + \ell_i )\no\\
		&\qquad\qquad\qquad - (\alpha_i + \beta_i + \gamma_i)c\l( e^{s_i}\l( \tfrac{1}{2} + \alpha_i \r) + \ell_i \r)\B c(e^{s_i}(\tfrac{1}{2} + \alpha_i + \beta_i) + \ell_i ) \B \Big\}\no\\
                &= \prod_{i=1}^{k}\Big\{ \left( \frac{1}{2} + \alpha_i \right)\B c(e^{s_i}(\tfrac{1}{2} + \alpha_i + \beta_i) + \ell_i ) + \left( \frac{1}{2} + \alpha_i + \beta_i \right)c(e^{s_i}(\tfrac{1}{2} + \alpha_i) + \ell_i)\B \no\\
                &\qquad\qquad\qquad\qquad\qquad\qquad\qquad\qquad\qquad\qquad\qquad - (\alpha_i + \beta_i + \gamma_i)c\l( e^{s_i}\l( \tfrac{1}{2} + \alpha_i \r) + \ell_i \r)\B \Big\}\no\\
                &=\prod_{i=1}^{k}\left\{ \left( \frac{1}{2} + \alpha_i \right) \B c(e^{s_i}(\tfrac{1}{2} + \alpha_i + \beta_i) + \ell_i ) - \l( \frac{1}{2} - \gamma_i \r)\B c\l( e^{s_i}\l( \tfrac{1}{2} + \alpha_i \r) + \ell_i \r) + \l( \frac{1}{2} - \gamma_i \r) \right\}.\label{ghostoperator}
	\end{align}
These operators are able to be calculated explicitly in the $k$-th order boundary state. This quantity is given in (\ref{ghost_op_calc}). 
Assuming that all $\alpha_i,\;\beta_i,\;\gamma_i $ integrals of (\ref{ghost_op_calc}) converge and we obtain the $k$-th order boundary state;
	\begin{align}
        	&|B_*^{(k)}(\Psi)\rangle \no\\
                &=  \frac{s^{k}}{k!}\Bigg( \frac{e^s}{ e^s - 1}\prod_{i=1}^{k}\int_0^\infty d\alpha_i f(\alpha_i) \int_0^\infty d\beta_i \tilde f(\beta_i) \int_0^\infty d\gamma_i f(\gamma_i)  \left( \frac{1}{2} + \alpha_i \right)\no\\
                &\qquad\qquad\qquad\qquad  -  \frac{1}{e^s - 1}  \prod_{i=1}^{k}\int_0^\infty d\alpha_i f(\alpha_i) \int_0^\infty d\beta_i \tilde f(\beta_i) \int_0^\infty d\gamma_i f(\gamma_i) \l( \frac{1}{2} - \gamma_i \r)\Bigg)  |B\rangle\no\\
                &= \frac{s^k}{k!} \frac{e^s}{e^s - 1} \prod_{i=1}^{k}\left[  \left( \frac{1}{2}F(0) + \del F(0) \right)\tilde F(0) F(0) \right]|B\rangle\no\\
                &\qquad\qquad\qquad\qquad\qquad\qquad - \frac{s^k}{k!}\frac{1}{e^s - 1}\prod_{i=1}^{k}  \left[\left( \frac{1}{2}F(0) - \del F(0) \right)\tilde F(0) F(0) \right]|B\rangle \no\\
                &= \frac{s^k}{k!}\frac{e^s}{e^s - 1}\left[\frac{1}{2}(F^2(0) +  \del F^2(0) )\tilde F(0)\right]^k | B \rangle - \frac{s^k}{k!}\frac{1}{e^s - 1}\left[\frac{1}{2}(F^2(0) -  \del F^2(0) )\tilde F(0)\right]^k | B \rangle \label{k_th_boundary},
        \end{align}
where we used
	\begin{align}
        	\int_0^sds_1 \int^s_{s_1}ds_2 \int^s_{s_2}ds_3 \cdots \int^s_{s_{k-1}}ds_k = \frac{s^k}{k!}.
        \end{align}
Therefore
	\begin{align}
        	&|B_*(\Psi)\rangle \no \\
                &= \sum_{k=0}^{\infty}\left[ \frac{s^k}{k!}\frac{e^s}{e^s - 1}\left[\frac{1}{2}(F^2(0) +  \del F^2(0) )\tilde F(0)\right]^k | B \rangle - \frac{s^k}{k!}\frac{1}{e^s - 1}\left[\frac{1}{2}(F^2(0) -  \del F^2(0) )\tilde F(0)\right]^k | B \rangle \right] \no\\
                &= \frac{e^s}{e^s - 1}\exp\left( \frac{s}{2}\left[(F^2(0) +  \del F^2(0) )\tilde F(0)\right] \right)|B\rangle  - \frac{1}{e^s - 1}\exp\left( \frac{s}{2}\left[(F^2(0) -  \del F^2(0) )\tilde F(0)\right] \right)|B\rangle \no \\
                &= \frac{e^s}{e^s - 1}\limz \exp\l( \frac{s}{2}z\frac{1-G(z) -G'(z)}{G(z)} \r)|B\rangle - \frac{1}{e^s - 1}\limz \exp\l( \frac{s}{2}z\frac{1-G(z) +G'(z)}{G(z)} \r)|B\rangle \label{boundaryexp},
	\end{align}
where $G = 1-F^2$. This relation is the boundary state for the class of analytic solutions (\ref{solution}).
\\\\
\textbf{$\bm{s\rightarrow 0}$ limit}\\
(\ref{k_th_boundary}) with $k=1$ becomes 
	\begin{align}
        	|B_*^{(1)}(\Psi)\rangle = \frac{se^s}{e^s - 1}\left[\frac{1}{2}(F^2(0) +  \del F^2(0) )\tilde F(0)\right] | B \rangle - \frac{s}{e^s - 1}\left[\frac{1}{2}(F^2(0) -  \del F^2(0) )\tilde F(0)\right] | B \rangle.
        \end{align}
In $s \rightarrow 0$, we obtain
	\begin{align}
        	|B_*^{(1)}(\Psi)\rangle 
                		  &\rightarrow \del F^2(0)\tilde F(0)| B \rangle \no\\
                                  &=\lim_{z\rightarrow 0}z\frac{\del F^2(z)}{1-F^2(z)}|B\rangle \no\\
                                  &=-\limz z\frac{G'(z)}{G(z)}|B\rangle \no.
        \end{align}
In what follows, all $z$ dependence on $G$'s and $F$'s is dropped. 
Therefore,
	\begin{align}
        	  |B\rangle + |B_*^{(1)}(\Psi)\rangle \underset{s\rightarrow 0}{\longrightarrow}  \left( 1-\limz z\frac{G'}{G} \right)|B\rangle.\label{k1limit}
        \end{align}
Moreover, for the boundary state (\ref{boundaryexp}),
	\begin{align}
        	|B_*(\Psi)\rangle
                &\underset{s\rightarrow 0}{\longrightarrow}\left[1 + \limz\left( z\frac{1-G-G'}{2G} - z\frac{1-G+G'}{2G}\right)\right]|B\rangle \no\\
                &= \left(1-\limz z\frac{G'}{G}\right)|B\rangle.\label{alllimit}
        \end{align}
Thus, in $s \rightarrow 0$ limit, our boundary state (\ref{alllimit}) is identified with the zeroth and the first order boundary state (\ref{k1limit}). This relation is consistent with the claim of \cite{Kiermaier:2008qu}.
\\\\
\textbf{Perturbative vacuum}\\
We know that the perturbative vacuum solution is obtained by selecting zero as $F(K)$, so $G(K) = 1$. In this case, (\ref{boundaryexp}) becomes
	\begin{align}
        	|B_*(\Psi_{\mathrm{per}}) \rangle &= \frac{e^s}{e^s - 1}\limz \exp\l( \frac{s}{2}z\frac{1-1 -0}{ 1} \r)|B\rangle - \frac{1}{e^s - 1}\limz \exp\l( \frac{s}{2}z\frac{1-1 +0}{1} \r)|B\rangle \no\\
                                                  &= |B\rangle. \label{pertursol_all}
        \end{align}
This is the expected result. Namely, the D-brane exists at the perturbative vacuum. Note that (\ref{boundaryexp}) loses the $s$ dependence for the perturbative solution. Moreover, if we substitute $G(K)=1$ into (\ref{k1limit}), we find
        \begin{align}
        	|B\rangle + |B_*^{(1)}(\Psi_{\mathrm{per}}) \rangle \rightarrow |B\rangle \label{pertursol_k_1}.
        \end{align}	
Therefore, the boundary state (\ref{pertursol_all}) coincides with (\ref{pertursol_k_1}) without a $Q$-exact term. This is consistent with (\ref{s_property}).

A pure gauge solution also describes the perturbative vacuum. This solution is obtained by changing $F(K)$ of (\ref{solution}) to $F(K+\epsilon)$, where $\epsilon$ is a parameter. For example,
	\begin{align}
        	F(K+\epsilon) = \frac{1}{\sqrt{1-(K+\epsilon)}},
        \end{align}
or
	\begin{align}
        	G(K+\epsilon) = 1 - F^2(K+\epsilon) = - \frac{K+\epsilon}{1-(K+\epsilon)} \label{puregauge}
        \end{align}
gives the pure gauge solution $\Psi_{\mathrm{pure}}$. In this case, from (\ref{alllimit}), the $s\rightarrow 0$ boundary state becomes 
	\begin{align}
        	|B_*(\Psi_{\mathrm{pure}})\rangle &\rightarrow \left(1 - \limz z\left( -\frac{1}{(1-(z+\epsilon))^2} \right)\left( -\frac{1-(z+\epsilon)}{z+\epsilon} \right)\right)|B\rangle\no\\
                				  &=|B\rangle.
        \end{align}
Moreover, we substitute (\ref{puregauge}) into (\ref{boundaryexp});
	\begin{align}
        	|B_*(\Psi_{\mathrm{pure}})\rangle &= \frac{e^s}{e^s-1}\limz \left( \frac{s}{2}z\left[ 1 + \frac{z+\epsilon}{1-(z+\epsilon) }+ \frac{1}{(1-(z+\epsilon))^2} \right]\left[ -\frac{1-(z+\epsilon) }{z+\epsilon} \right] \right)|B\rangle \no\\
                				  &\qquad - \frac{1}{e^s-1}\limz \left( \frac{s}{2}z\left[ 1 + \frac{z+\epsilon}{1-(z+\epsilon) }- \frac{1}{(1-(z+\epsilon))^2} \right]\left[ -\frac{1-(z+\epsilon) }{z+\epsilon} \right] \right)|B\rangle \no\\
                                                  &= |B\rangle.
        \end{align}
We obtain the expected boundary state in $s\rightarrow 0$ and all order for the pure gauge solution.
\\\\
\textbf{Tachyon vacuum}\\
When $F(K) = \frac{1}{\sqrt{1-K}}$, so $G(K) = -\frac{K}{1-K}$, we know that the tachyon vacuum solution is obtained. From (\ref{boundaryexp}), 
		\begin{align}
        	|B_*(\Psi_{\mathrm{tach}})\rangle &= \frac{e^s}{e^s - 1}\limz \exp \left[\frac{sz}{2}\l( 1 + \frac{z}{1-z} + \frac{1}{(1-z)^2}  \r)\l( - \frac{z}{1-z} \r)^{-1}  \right]|B\rangle \no\\
                				  &\qquad - \frac{1}{e^s - 1}\limz \exp \left[\frac{sz}{2}\l( 1 + \frac{z}{1-z} - \frac{1}{(1-z)^2}  \r)\l( - \frac{z}{1-z} \r)^{-1}  \right]|B\rangle \no\\
                                                  &= 0.\label{tachyon_boundary_state_all}
		\end{align}
This is consistent with the fact that the D-brane disappears at the tachyon vacuum. And from (\ref{k1limit}),
        \begin{align}
        	|B\rangle + |B_*^{(1)}(\Psi_{\mathrm{tach}})\rangle &\rightarrow \left( 1  -\limz z\left( -\frac{1}{(1-z)^2} \right) \l(  -\frac{z}{1-z} \r)^{-1}\r)|B\rangle \no\\
                                  &=0.\label{tachyon_boundary_state_limit}
        \end{align}
(\ref{tachyon_boundary_state_all}) agrees with (\ref{tachyon_boundary_state_limit}) without a $Q$-exact term. This result is the same as one of \cite{Kiermaier:2008qu}.

\section{Comments on multi-brane and ghost brane solutions}
Recently, Murata and Schnabl suggested the possible presence of multi-brane solutions \cite{Murata:2011ex}. They calculated the energy of (\ref{solution}) and the result was
	\begin{align}
        	E = -\frac{1}{2\pi^2} \lim_{z\rightarrow 0} z\frac{G'}{G}.
        \end{align}
 If the function behave $G(z)$ as $z^{1-n}$, the energy is 
 	\begin{align}
        	E = \frac{-1+n}{2\pi^2}. \label{multi_energy}
        \end{align}
The energy at the tachyon vacuum is $-\frac{1}{2\pi^2}$ and that at the perturbative vacuum is zero. As the D-brane does not exist at the tachyon vacuum and exists at the perturbative vacuum, the value $\frac{1}{2\pi^2}$ is the tension of the D-brane. So, (\ref{multi_energy}) is the energy for configurations of $n$ D-branes.

When $G(z)\sim z^{1-n}$, our boundary state with $s \rightarrow 0$ (\ref{alllimit}) is
 	\begin{align}
        	|B_*(\Psi)\rangle = n|B\rangle.\label{multi_boundary_state}
        \end{align}
At the tachyon vacuum, $|B_*(\Psi_{\mathrm{tach}})\rangle = 0$, while at the perturbative vacuum, the boundary state $|B\rangle$ exists. So (\ref{multi_boundary_state}) means the configuration of $n$ D-branes too. Therefore our boundary state is similar to the result of \cite{Murata:2011ex}.
 It seems that $G(z)\sim z^{1-n}$ gives a solution that would describe the configuration of $n$ D-branes. But there is a problem. When $n \geq 2$,  $G(z)$ is singular for $z \rightarrow 0$. In other words, the integral value
 	\begin{align}
        	F(0) = \int^\infty_0 dt f(t) 
        \end{align}
 diverges. For example, when $n=2$, $G(z) = z^{-1}$, so $F(z)=\sqrt{1-\frac{1}{z}}$. But an integral value $F(0)$ diverges. Therefore we must regularize $F(z)$ to obtain a finite value. But even if we can obtain a finite value, our boundary state (\ref{boundaryexp}) has $s$ dependence in $n\geq 2$. Recall that the general boundary state (\ref{general_boundary_state}) is constructed by using a classical solution of the equation of motion and that it has the property that it changes only by a $Q$-exact term under the variation of $s$ due to the equation of motion. But $|B_*(\Psi)\rangle$ that ghost operators are calculated can not have a $Q$-exact term. In this case, if $|B_*(\Psi)\rangle$ has a $s$ dependence, a string field $\Psi$ is not a solution. (\ref{boundaryexp}) is in the instance. Thus even if we can regularize $F(K)$, we can not obtain multi-brane solutions from (\ref{solution}) at least naively
 
While, in $n\leq -1$, $G(z)\sim z^{1-n}$ is regular for $z\rightarrow 0$. In this case the energy and the $s\rightarrow 0$ limit boundary state are
	\begin{align}
        	E &= -\frac{|n-1|}{2\pi^2},\\
                |B_*(\Psi)\rangle &\rightarrow -|n||B\rangle,
        \end{align}
respectively. These mean that a ghost brane which has a negative tension exists. But our boundary state (\ref{boundaryexp}) becomes 
	\begin{align}
        	&|B_*(\Psi)\rangle\no\\
                &= \frac{e^s}{e^s - 1}\limz\exp\l( \frac{s}{2}(z^n-z-(1-n)) \r)|B\rangle - \frac{1}{e^s - 1}\limz\exp\l( \frac{s}{2}(z^n-z+(1-n)) \r)|B\rangle.\no\\
                &= \limz \exp\left( \frac{s}{2}(z^{-|n|}-z) \right)\frac{\sinh \frac{ns}{2}}{\sinh \frac{s}{2}}|B\rangle.
        \end{align}
Thus ghost brane solutions do not exist at least in the form (\ref{solution}) because this boundary state diverges for $z\rightarrow 0$\footnote{The author thanks Y Okawa and M. Murata for helpful comments. }.

\section{Conclusions and discussions}

We construct the boundary state for the class of analytic solutions of the open string field theory. It gives expected one which is for the perturbative vacuum solution and the tachyon vacuum solution. The result is that in $s\rightarrow 0$ limit, the boundary state of the zeroth and the first order coincides with one of all order. Our boundary state is constructed under the assumption that $F(0)$ is finite. In fact, it is the case for the perturbative and the tachyon vacuum solutions, but it is not for the multi-brane solutions. Thus we think that $F(K)$ must be regularized. But even if it can be done so, a string field $\Psi$ that obtained does not satisfy the equation of motion at least naively, because $|B_*(\Psi)\rangle$ has the $s$ dependence. If multi-brane solutions exist, the string fields (\ref{solution}) must be modified. Moreover ghost brane solutions do not exist at least in the form (\ref{solution}) because our boundary state diverges.

We calculate the boundary state for the solution in $KBc$ subalgebra. But one will also be obtained for solutions in extended subalgebras (for example \cite{Noumi:2011kn}\cite{Bonora:2011ns}) because the boundary state $|B_*(\Psi)\rangle$ consists of a general solution $\Psi$. Of course a boundary state for those solutions must also satisfy properties of one. So we think that it is convenient to find any solutions that we use this properties.

\section*{Acknowledgments}

The author thanks Tsunehide Kuroki, Toshifumi Noumi, Masaki Murata and Yuji Okawa for useful comments. The author also thanks the Yukawa Institute for Theoretical Physics at Kyoto University, where this work was initiated during the YITP-W-11-10 on "YONUPA Summer School 2011". 

\appendix
\section{Calculation of ghost operators}
In this appendix, We first prove 
	\begin{align}
        	&\prod_{i=1}^{m}\left[  A_i \B c(x_i) - D_i \B c(y_i) + D_i \right] \no\\
                &=  \prod_{i=1}^{m} A_i \B c(x_m) - \sum_{j=1}^{m}\Big\{\prod_{i=1}^{j}A_i\prod_{r=j+1}^{m}D_r\l( \B c(y_{j+1}) - \B c(x_{j}) \r)\Big\}  - \prod_{i=1}^m D_i (\B c(y_1) - 1)\label{appendix1}
        \end{align}
by induction, where $A_i$ and $D_i$ are constants. When $m=1$, the second term of the right hand side is not defined. So (\ref{appendix1}) holds. When $m=2$,
	\begin{align}
        	&\prod_{i=1}^{2}\left[  A_i \B c(x_i) - D_i \B c(y_i) + D_i \right]\no \\
                &= \left[  A_1 \B c(x_1) - D_1 \B c(y_1) + D_1 \right]\left[  A_2 \B c(x_2) - D_2 \B c(y_2) + D_2 \right]\no\\
                &= A_1 \B c(x_1)\left[  A_2 \B c(x_2) - D_2 \B c(y_2) + D_2 \right]\no\\
                &\qquad - D_1 \B c(y_1)\left[  A_2 \B c(x_2) - D_2 \B c(y_2) + D_2 \right]\no\\
                &\qquad\qquad + D_1 \left[  A_2 \B c(x_2) - D_2 \B c(y_2) + D_2 \right]\no\\
                &=   A_1A_2 \B c(x_2) - A_1D_2 \B c(y_2) + A_1D_2\B c(x_1)\no\\
                &\qquad  - A_2D_1 \B c(x_2) + D_1D_2 \B c(y_2) - D_1D_2\B c(y_1)\no \\
                &\qquad\qquad  +A_2D_1 \B c(x_2) - D_1D_2 \B c(y_2) + D_1D_2 \no\\
                &=   A_1A_2 \B c(x_2) - A_1D_2 \B c(y_2) + A_1D_2\B c(x_1) - D_1D_2\B c(y_1) + D_1D_2.\no
        \end{align}
Thus, in this case, (\ref{appendix1}) is right. (\ref{appendix1}) with $m = k+1$,
	\begin{align}
        	&\prod_{i=1}^{k+1}\left[  A_i \B c(x_i) - D_i \B c(y_i) + D_i \right] \no\\
                &=  \left[\prod_{i=1}^{k} A_i \B c(x_k) - \sum_{j=1}^{k-1}\Big\{\prod_{i=1}^{j}A_i\prod_{r=j+1}^{k}D_r\l( \B c(y_{j+1}) - \B c(x_{j}) \r)\Big\}  - \prod_{i=1}^k D_i (\B c(y_1) - 1)\right]\no\\
                &\qquad\qquad\qquad\qquad\qquad\qquad\qquad\qquad\qquad\qquad\qquad\qquad \times \left[  A_{k+1} \B c(x_{k+1}) - D_{k+1} \B c(y_{k+1}) + D_{k+1} \right] ,\no
        \end{align}
by using $\B c(t_1)\B c(t_2)=\B c(t_2)$ which follows $\{\B , c(t)\}=1$,
	\begin{align}
		&\sum_{j=1}^{k-1}\Big\{\prod_{i=1}^{j}A_i\prod_{r=j+1}^{k}D_r\l( \B c(y_{j+1}) - \B c(x_{j}) \r)\Big\}\B c(t)=0, \no\\
                &\prod_{i=1}^k D_i (\B c(y_1) - 1)\B c(t) = 0,\no
	\end{align}
and therefore
        \begin{align}
        	&\prod_{i=1}^{k+1}\left[  A_i \B c(x_i) - D_i \B c(y_i) + D_i \right] \no\\
        	&=\prod_{i=1}^{k} A_i \B c(x_k) \left[  A_{k+1} \B c(x_{k+1}) - D_{k+1} \B c(y_{k+1}) + D_{k+1} \right] \no\\
                  &\qquad - \sum_{j=1}^{k-1}\Big\{\prod_{i=1}^{j}A_i\prod_{r=j+1}^{k}D_r\l( \B c(y_{j+1}) - \B c(x_{j}) \r)\Big\}D_{k+1}\no\\
                  &\qquad\qquad  - \prod_{i=1}^k D_i (\B c(y_1) - 1)  D_{k+1}  \no\\
                &=\prod_{i=1}^{k+1} A_i \B c(x_{k+1})  - \prod_{i=1}^{k}A_iD_{k+1} \B c(y_{k+1}) +\prod_{i=1}^{k}A_i D_{k+1}\B c(x_k) \no\\
                  &\qquad - \sum_{j=1}^{k-1}\Big\{\prod_{i=1}^{j}A_i\prod_{r=j+1}^{k+1}D_r\l( \B c(y_{j+1}) - \B c(x_{j}) \r)\Big\}\no\\
                  &\qquad\qquad  - \prod_{i=1}^{k+1} D_i(\B c(y_1)-1) \no\\
                &=\prod_{i=1}^{k+1} A_i \B c(x_{k+1}) - \sum_{j=1}^{k}\Big\{\prod_{i=1}^{j}A_i\prod_{r=j+1}^{k+1}D_r\l( \B c(y_{j+1}) - \B c(x_{j}) \r)\Big\} - \prod_{i=1}^{k+1} D_i(\B c(y_1)-1).\no
        \end{align}
Therefore (\ref{appendix1}) with $m=k+1$ holds. Moreover when we calculate the boundary state, we can use (\ref{formula2}). By using this, because $\B c(t)$ is the constant, the second term in the above equation vanishes. Thus when we calculate the boundary state, we substitute $A_i=\frac{1}{2} + \alpha_i$, $D_i =  \frac{1}{2} - \gamma_i$ and $\B c(t) = \frac{e^s}{e^s - 1}$ into (\ref{appendix1}), the ghost operators become 
	\begin{align}
        	&\prod_{i=1}^{k}\left\{ \left( \frac{1}{2} + \alpha_i \right) \B c(e^{s_i}(\tfrac{1}{2} + \alpha_i + \beta_i) + \ell_i ) - \l( \frac{1}{2} - \gamma_i \r)\B c\l( e^{s_i}\l( \tfrac{1}{2} + \alpha_i \r) + \ell_i \r) + \l( \frac{1}{2} - \gamma_i \r) \right\}\no\\
                &= \frac{e^s}{e^s - 1}\prod_i^k \l( \frac{1}{2} + \alpha_i\r) - \frac{1}{e^s - 1}\prod_i^k \l( \frac{1}{2} - \gamma_i\r)\label{ghost_op_calc}.
        \end{align}


\begin{thebibliography}{99} 

\bibitem{Witten:1985cc}
  E.~Witten,
  ``Noncommutative Geometry and String Field Theory,''
  Nucl.\ Phys.\  B {\bf 268}, 253 (1986).

\bibitem{Schnabl:2005gv}
  M.~Schnabl,
  ``Analytic solution for tachyon condensation in open string field theory,''
  Adv.\ Theor.\ Math.\ Phys.\  {\bf 10}, 433 (2006)
  [arXiv:hep-th/0511286].
\bibitem{Okawa:2006sn}
  Y.~Okawa, L.~Rastelli and B.~Zwiebach,
  ``Analytic Solutions for Tachyon Condensation with General Projectors,''
  arXiv:hep-th/0611110.

\bibitem{Schnabl:2007az}
  M.~Schnabl,
  ``Comments on marginal deformations in open string field theory,''
  Phys.\ Lett.\  B {\bf 654}, 194 (2007)
  [arXiv:hep-th/0701248].

\bibitem{Erler:2007rh}
  T.~Erler,
  ``Marginal Solutions for the Superstring,''
  JHEP {\bf 0707}, 050 (2007)
  [arXiv:0704.0930 [hep-th]].

\bibitem{Okawa:2007ri}
  Y.~Okawa,
  ``Analytic solutions for marginal deformations in open superstring field theory,''
  JHEP {\bf 0709}, 084 (2007)
  [arXiv:0704.0936 [hep-th]].

\bibitem{Kiermaier:2007vu}
  M.~Kiermaier and Y.~Okawa,
  ``Exact marginality in open string field theory: A General framework,''
  JHEP {\bf 0911}, 041 (2009)
  [arXiv:0707.4472 [hep-th]].

\bibitem{Kiermaier:2007ki}
  M.~Kiermaier and Y.~Okawa,
  ``General marginal deformations in open superstring field theory,''
  JHEP {\bf 0911}, 042 (2009)
  [arXiv:0708.3394 [hep-th]].
  
\bibitem{Okawa:2006vm}
  Y.~Okawa,
  ``Comments on Schnabl's analytic solution for tachyon condensation in
  Witten's open string field theory,''
  JHEP {\bf 0604}, 055 (2006)
  [arXiv:hep-th/0603159].


\bibitem{Erler:2009uj}
  T.~Erler and M.~Schnabl,
  ``A Simple Analytic Solution for Tachyon Condensation,''
  JHEP {\bf 0910}, 066 (2009)
  [arXiv:0906.0979 [hep-th]].




\bibitem{Kiermaier:2010cf}
  M.~Kiermaier, Y.~Okawa and P.~Soler,
  ``Solutions from boundary condition changing operators in open string field theory,''
  JHEP {\bf 1103}, 122 (2011)
  [arXiv:1009.6185 [hep-th]].

\bibitem{Bonora:2010hi}
  L.~Bonora, C.~Maccaferri and D.~D.~Tolla,
  ``Relevant Deformations in Open String Field Theory: a Simple Solution for Lumps,''
  arXiv:1009.4158 [hep-th].

\bibitem{Bonora:2011ri}
  L.~Bonora, S.~Giaccari and D.~D.~Tolla,
  ``The energy of the analytic lump solution in SFT,''
  JHEP {\bf 1108}, 158 (2011)
  [arXiv:1105.5926 [hep-th]].

\bibitem{Erler:2011tc}
  T.~Erler and C.~Maccaferri,
  ``Comments on Lumps from RG flows,''
  arXiv:1105.6057 [hep-th].

\bibitem{Maccaferri:2011zz}
  C.~Maccaferri,
  ``A solution for relevant deformations in open string field theory,''
  Prog.\ Theor.\ Phys.\ Suppl.\  {\bf 188}, 83 (2011).

\bibitem{Noumi:2011kn}
  T.~Noumi and Y.~Okawa,
  ``Solutions from boundary condition changing operators in open superstring field theory,''
  arXiv:1108.5317 [hep-th].
  
\bibitem{Bonora:2011ns}
  L.~Bonora, S.~Giaccari, D.~D.~Tolla,
  ``Lump solutions in SFT. Complements,''
  [arXiv:1109.4336 [hep-th]].

\bibitem{Murata:2011ex}
  M.~Murata and M.~Schnabl,
  ``On Multibrane Solutions in Open String Field Theory,''
  Prog.\ Theor.\ Phys.\ Suppl.\  {\bf 188}, 50 (2011)
  [arXiv:1103.1382 [hep-th]].





\bibitem{Kiermaier:2008qu}
  M.~Kiermaier, Y.~Okawa and B.~Zwiebach,
  ``The boundary state from open string fields,''
  arXiv:0810.1737 [hep-th].


\bibitem{Ellwood:2008jh}
  I.~Ellwood,
  ``The Closed string tadpole in open string field theory,''
  JHEP {\bf 0808}, 063 (2008)
  [arXiv:0804.1131 [hep-th]].

\bibitem{Okuda:2006fb}
  T.~Okuda and T.~Takayanagi,
  ``Ghost D-branes,''
  JHEP {\bf 0603}, 062 (2006)
  [arXiv:hep-th/0601024].

\bibitem{Erler:2006ww}
  T.~Erler,
  ``Split String Formalism and the Closed String Vacuum, II,''
  JHEP {\bf 0705}, 084 (2007).
  [hep-th/0612050].
  
  
\bibitem{Kiermaier:2007jg}
  M.~Kiermaier, A.~Sen and B.~Zwiebach,
  ``Linear b-Gauges for Open String Fields,''
  JHEP {\bf 0803}, 050 (2008)
  [arXiv:0712.0627 [hep-th]].


\bibitem{Kiermaier:2008jy}
  M.~Kiermaier and B.~Zwiebach,
  ``One-Loop Riemann Surfaces in Schnabl Gauge,''
  JHEP {\bf 0807}, 063 (2008)
  [arXiv:0805.3701 [hep-th]].

\end{thebibliography}
\end{document}